\documentclass[twoside,slac_one]{revtex4}
\usepackage{graphicx}
\usepackage{fancyhdr}
\usepackage{amsmath} 
\usepackage{bm}
\usepackage{amsxtra}
\usepackage{amssymb}
\usepackage{amsthm}
\usepackage{latexsym}
\usepackage{lscape}
\usepackage{xspace}

\newcommand{\mymet}{\makebox[2.4ex]{\ensuremath{\not\!\! E_{\mathrm{T}}}}}
\newcommand{\two}{2D-analysis\xspace}
\newcommand{\bdt}{BDT-analysis\xspace}
\providecommand{\tev}{\ensuremath{\mathrm{\;TeV}}\xspace}
\providecommand{\gevc}{\ensuremath{\mathrm{\;GeV}/c}\xspace}
\providecommand{\gevcc}{\ensuremath{\mathrm{\;GeV}/c^2}\xspace}
\providecommand{\pbinv}{\ensuremath{\mathrm{pb}^{-1}}}
\providecommand{\pt}{\ensuremath{p_{\rm T}}}
\providecommand{\ttbar} {\ensuremath{t \bar t}\xspace}

\begin{document}

\title{Single Top production at CMS}

%

\author{T. Speer}
\affiliation{Department of Physics and Astronomy, Brown University, Providence, RI, USA}

\begin{abstract}
A first measurement of the cross section of single top quark production in the $t$ channel in pp collision at $\sqrt{s}=7$ TeV is presented. The measurement is performed on a data sample corresponding to an integrated luminosity of 35.9 pb$^{-1}$ recorded at the LHC with the CMS detector.
Leptonic decay channels with an electron or a muon in the final state are considered. After a selection optimized for the $t$-channel mode,
two different and complementary analyses have been performed.
 Both analyses confirm the Tevatron's observation of single top, and their combination measures a cross section of $\sigma = 83.6 \pm 29.8~\mathrm{(stat.+syst.)} \pm 3.3~\mathrm{(lumi.)\,\,pb}$ which is consistent with the Standard Model prediction.
\end{abstract}

\maketitle

\thispagestyle{fancy}


\section{Introduction}

The existence of  single top production has been established by the D0 and CDF experiments at the Tevatron $p\bar p$ collider~\cite{Abazov:2009ii,Aaltonen:2009jj,Group:2009qk} and the first measurements of individual production mechanisms have recently appeared~\cite{Abazov:2009pa,Aaltonen:2010jr}.
Three different production mechanisms are foreseen in the Standard Model (SM) $t$ channel, $s$ channel, and $tW$ (or $W$-associated).
In 7~TeV proton-proton collisions the $t$-channel mode, is by far the most abundant of the three mechanisms and it is the one with the most striking final state topology. Next-to-leading order (NLO) computations predict $\sigma_{t,4FS}=59.1^{+3.0}_{-4.0}$~pb in the 4-flavour scheme and $\sigma_{t,5FS}=62.3^{+2.3}_{-2.4}$~pb in the 5-flavour~\cite{Campbell:2009gj}, for a top mass of $m_t=172.5\,\mbox{GeV/c}^2$.

We present here first evidence for $t$-channel single top quark production in $pp$ collisions at $\sqrt{s}=7\tev$ at the LHC, with a first measurement of the production cross section~\cite{CMS-PAS-TOP-10-008}.
The results are based on the data sample sample collected in 2010 by the
 Compact Muon Solenoid (CMS) experiment~\cite{JINST}, and corresponds to
 an integrated luminosity of $35.9\pm1.4~\pbinv$.

This measurement is performed in the leptonic decay channel, in which the W
boson decays into an electron or a muon.
The $t$-channel production mode is treated as signal and the other two
 production modes will be considered as background.

After a dedicated event selection, two complementary analyses are performed.
In the first analysis, referred to as the 2D-analysis, a data driven method using two angular properties specific to $t$-channel top quark production will be used.
In the second analysis, referred to as the BDT-analysis, the overall compatibility of the signal event
 candidates with the Standard Model expectations of electroweak top quark
 production is probed by using a multivariate analysis technique.


\section{The event selection}

Both analyses employ similar reconstruction techniques and selection criteria.
 Signal events are characterized by exactly one isolated muon or electron and missing transverse energy from  the leptonic decay of the $W$ boson as well as by one central $b$-jet from the top quark decay and an additional light-quark jet from the hard scattering process. The latter is found most often in the forward direction.

Events are required to pass either a single electron trigger or a single
 muon trigger.
The minimum $E_{\rm T}$ requirement for the electron trigger ranged from 10
 GeV to 22 GeV, while the minimum $p_{\rm T}$ requirement for the muon
 trigger ranged from 9 GeV to 15 GeV.
The selected data sample is used both for the selection of the signal and for signal-depleted control regions used for data-driven background studies. Therefore no lepton isolation criteria were used at trigger level, in order to allow background estimations based on samples failing these criteria.

After offline reconstruction, events are selected requiring exactly one isolated lepton (electron or muon) with $\pt>20\gevc$ and $|\eta|<2.4$ for muons and $\pt > 30\gevc$, $| \eta | < 2.5$ for electrons, and
exactly two jets ($\pt > 30\gevc$, $| \eta | < 5$).
In the 2D analysis, only the jets and the missing transverse energy
 ($\mymet$) are reconstructed with the particle flow
 algorithm~\cite{particleflow}, while in the BDT-analysis, all objects are
 reconstructed with the particle flow algorithm.

In order to reduce the large background from $W+$ light partons, one of the two selected jets is required to be identified as a b-jet according to the tight selection criteria of the {\em track counting} (TC) $b$-tagging algorithm~\cite{btag}.
To further reduce the background, the \two requires that the second jet is
 not tagged according to a looser selection criteria, since most of the
 signal events are expected to have only one $b$ quark inside the acceptance
 of the tracking detectors ($|\eta| < 2.5$).
The \bdt does not make this requirement in order to increase statistics and  profit from the larger separation power of the BDT discriminant.

One the other hand, to remove the kinematic region where
 the two identified jets are back-to-back, the \bdt requires the two selected jets to satisfy
 $\Delta\phi(j_1,j_2)<3.0$.
This region is found to be poorly reproduced by the simulation in a sample
 enriched in $W+$light partons, affecting some of the observables used in
 the analysis.

Finally, to further suppress contributions from processes where the lepton does not come from a leptonically decaying $W$ boson, the transverse mass is required to be $M_T >40\gevcc$ for muon events and $M_T >50\gevcc$ for electron events.
The transverse mass is defined as
\begin{equation}
M_{T} = \sqrt{\left(p_{T,l} + p_{T,\nu}\right)^2
			- \left( p_{x,l} + p_{x,\nu} \right)^2
			- \left(  p_{y,l} + p_{y,\nu} \right)^2}~,
\label{eqn:mTW}
\end{equation}
where the neutrino momentum vector is assumed equal to the missing transverse energy ($\mymet$).

In data, the number of selected events is 72 in the electron and 112 in the muon channels in the \two, and 82 in the electron and 139 in the muon channels in the \bdt.

\subsection{Top quark reconstruction}

Both analyses require the reconstruction of the 4-momentum of the top-quark candidate.
A constrained kinematic fit is used to reconstruct the complete kinematics
 of the event under the hypothesis that it is a single top event decaying
 into a lepton+jets final state.
This leads to a quadratic equation in the longitudinal neutrino momentum, $p_{z,\nu}$.
Solutions to this equation can have an imaginary part when $M_T$ is larger than the $W$ pole mass used in the constraint.
The imaginary component is then eliminated by modifying the \mymet\ such as to give $M_T = M_W$, still respecting the $W$ mass constraint.
 When two real solutions are present, which happens in 77.6\% of the cases, the solution with the smallest $|p_{z,\nu}|$ is chosen, which, in simulated events, is correct in 60.3\% of the cases.

Choosing the jet with the highest $b$-tagging discriminant as the jet originating in the decay of the top quark is correct in simulated events in 92.6\% (87.4\%) of the cases after the 2D (BDT) selection.
 The non-tagged jet is matched to the recoil quark in 89.6\% (84.0\%) of the cases.

\section{Background Estimation}

With a relatively small signal and a large background, one crucial element
 of the analysis is the determination of the background. 
Data-driven methods are thus used to estimate the main backgrounds.
The event yields for the two selections are summarized in Table~\ref{tab:evtYield_with_datadriven}.

\subsection{QCD estimation}

This analysis probes a very specific kinematical phase space populated only
 by the tails of the QCD distributions.
This, despite the excellent agreement of the CMS simulation with data, together with the small number of selected events in the simulation, makes the estimate of this background on simulated events not very significant.

The normalization of this background is estimated by a profile likelihood fit to the $M_T$ distribution after all other selection criteria have been applied, by parametrizing the $M_T$ distribution as
$F(M_T)= a\cdot S(M_T)+b\cdot B(M_T)$,
 where $S(M_T)$ and $B(M_T)$ are templates for signal-like (leptons coming from $W$ decays) and QCD-like events, respectively. 
 The $S(M_T)$ template is taken from the simulation, while the $B(M_T)$ template is extracted from a high statistics background dominated data sample composed mainly of QCD events.
This sample is obtained by removing the $b$-tagging and $M_T$ requirements and
 inverting the isolation cut. 
These requirements reject most of the signal-like events (single top,
 $W+X$, \ttbar, and in general any process with a charged lepton from an
 intermediate $W$ boson).

This procedure yields the following predictions for the number of QCD events passing the $M_T$ threshold in the 2D analysis:
\begin{align}
\label{eq:fitresultsQCD_m_2d}
N_{qcd}^{2D} & =  0.62 \pm 0.12 ~\mathrm{(stat.)}  \pm 0.08 ~\mathrm{(shape)}\pm 0.15 ~\mathrm{(stability)} \tag{muons}  \\
\label{eq:fitresultsQCD_e_2d}
N_{qcd}^{2D} & =  2.6  \pm 0.6  ~\mathrm{(stat.)}  \pm 3.1 ~\mathrm{(shape)} \pm 1.2  ~\mathrm{(stability)} \tag{electrons}  
\end{align}
and in the BDT analysis:
\begin{align}
\label{eq:fitresultsQCD_m_bdt}
N_{qcd}^{BDT} & =  4.92 \pm 0.99 ~\mathrm{(stat.)}  \pm 0.05~\mathrm{ (shape)}\pm 0.81 ~\mathrm{(stability)} \tag{muons}  \\
\label{eq:fitresultsQCD_e_bdt}
N_{qcd}^{BDT} & =  5.27 \pm 1.24 ~\mathrm{(stat.)}  \pm 0.79 ~\mathrm{(shape)} \pm 3.23 ~\mathrm{(stability)}   \tag{electrons}  
\end{align}
where ``$shape$'' indicates the systematic uncertainty coming from the $B(M_T)$ model and 
 ``$stability$'' indicates the maximum variation between the results when varying the fit range.
The central values of these predictions are used in the analyses, while the uncertainties on these values are conservatively taken as 
 $\pm 50\%$ in the muon decay channel in both analyses, $\pm 100\%$ in the electron decay channel in the BDT analysis and $^{+130\%}_{-100\%}$ in the 2D analysis.

\subsection{$W+$light partons estimation}

The $W+$light partons background is treated differently in both analyses.
In the BDT analysis, the $W+$light partons yield is treated as a nuisance
 parameter in a fully Bayesian procedure.
In the 2D analysis, partially data-driven methods are used to extract the
 normalization and the kinematics.
The same factor is then also used for $Z +$ jets. 

A suitable control sample, dominated by the $W$ + light flavors background, is obtained with an orthogonal selection where the events are required to have one isolated lepton and exactly two jets. One of the jets is required to be ``taggable'', i.e., within the tracker acceptance and with at least two tracks satisfying the quality selection of the $b$-tagging algorithm. Both jets should fail the tight $b$-tagging selection.
To model the distributions of the variables used in this analysis  in $W$ + light flavor background events in the signal region, the distributions obtained in this $W-$enriched sample in data will be used, after subtracting the other contributions (including signal, which accounts for roughly 1\% of this sample) estimated with simulated samples.


To estimate the scale factor for the $W+$light partons background components
both this $W$-enriched control sample (control sample $A$) and  a subset where at least one jet passes the loose $b$-tagging selection (but fails the tight one -- control sample $B$) are used.
In both samples a fit on the full $M_T$ distribution is performed. The QCD and $W+$light partons components are free parameters in the fit, while all other processes, including the heavy flavour components and the $t$-channel signal, are constrained to the expected values.
The scale factors between the number of events in the $W$-enriched control
 sample in data and simulation found are given in Table~\ref{tab:wlight-prediction}.

\begin{table}
\begin{center}
  \begin{tabular}{ |l|c|c| }
    \hline
    Process       & $SF$ from region $A$ & $SF$ from region $B$ \\
    \hline
    $\mu$ channel & $1.02\pm 0.03$ & $1.27\pm 0.09$ \\
    $e$ channel   & $0.97\pm 0.04$ & $1.05\pm 0.11$\\
\hline
\end{tabular}
\end{center}
\caption{Scale factors for $W+$light partons predicted by the fits in control regions $A$ and $B$ in the 2D analysis. Uncertainties are statistical only.}
\label{tab:wlight-prediction}
\end{table}

The 2D analysis takes as central predictions those from control sample $B$, upon the argument that it is closer to the signal region, obtaining an expectation of 18.2 (116) $W+$light parton events in the signal region for the muon (electron) decay channel.
An uncertainty of  $\pm 30\%$ ($\pm 20\%$) is then used, which covers both the statistical uncertainty from the fit and the difference between both predictions.
 The same scale factors are applied to $Z+$jets.

\subsection{Other background contributions}

The $VQ\bar Q$ and $Wc$ components are scaled to LO values and, on top of this correction, by further factors $2\pm 1$ and $1^{+1}_{-0.5}$, respectively, in order to take into account the results of the \ttbar cross section measurement exploiting $b$-tagging~\cite{top-10-003}, from which the \ttbar cross section itself is taken.
 The theory prediction is used for $VV$~\cite{Gavin:2010az} and single top in $s$~\cite{Kidonakis:2010tc} and $tW$~\cite{Campbell:2005bb} channels.
 The uncertainties on these values are considered as components of the systematic uncertainty.
 The BDT analysis treats the normalization of these backgrounds as a nuisance parameter in the fit, with Gaussian constraints corresponding to the systematic uncertainty.

\begin{table}
\begin{center}
\caption{Event yields summary, including data-driven estimations and $b$-tagging scale factors. The signal ($\dagger$) is normalized to the 5-flavour computation with the corresponding uncertainty~\cite{Campbell:2009gj}.}
\begin{tabular}{ |c||c|c||c|c| }
\hline
Process                      & 2D, $\mu$ channel          & 2D, $e$ channel            & BDT, $\mu$ channel         & BDT, $e$ channel\\
\hline \hline
single top, $t$ channel ($\dagger$)      & 17.6 $\pm$ 0.7  & 11.2 $\pm$ 0.4  & 17.6 $\pm$ 0.7  & 10.7 $\pm$ 0.5 \\ \hline
single top, $s$ channel      & 0.9 $\pm$ 0.3              & 0.6 $\pm$ 0.2              & 1.4 $\pm$ 0.5              & 1.0 $\pm$ 0.3 \\
single top, $tW$             & 3.1 $\pm$ 0.9              & 2.4 $\pm$ 0.7              & 3.8 $\pm$ 1.1              & $<$ 0.1  \\
$WW$                         & 0.29 $\pm$ 0.09            & 0.23 $\pm$ 0.07            & 0.32 $\pm$ 0.10            & 0.23 $\pm$ 0.07    \\
$WZ$                         & 0.24 $\pm$ 0.07            & 0.17 $\pm$ 0.05            & 0.33 $\pm$ 0.10            & 1.5 $\pm$ 0.4   \\
$ZZ$                         & 0.018$\pm$ 0.005           & 0.011 $\pm$ 0.003          & 0.020 $\pm$ 0.006          & $<$ 0.1  \\ \hline
$W+$ light partons           & 18.2 $\pm$ 5.5             & 11.6 $\pm$ 2.3             & 8.4 $\pm$ 4.2              & 7.0 $\pm$ 3.5  \\
$Z+X$                        & 1.7  $\pm$ 0.5             & 1.6  $\pm$ 0.3             & 0.7 $\pm$ 0.2              & 0.05 $\pm$ 0.03  \\
QCD                          & 0.6  $\pm$ 0.3             & 2.6$^{+3.4}_{-2.6}$        & 4.9 $\pm$ 2.5              & 5.3 $\pm$ 5.3  \\ \hline
$VQ\bar Q$                   & 20.4 $\pm$ 10.2            & 14.1 $\pm$ 7.1             & 17.6 $\pm$ 8.8             & 11.7 $\pm$ 5.8   \\
$Wc$                         & 12.9 $^{+12.9}_{-6.5}$     & 9.4 $^{+9.4}_{-4.7}$       & 9.2 $^{+9.2}_{-4.6}$       & 5.9 $^{+5.9}_{-2.9}$   \\
\ttbar                       & 20.3 $\pm$ 3.6             & 15.6 $\pm$ 2.8             & 34.9 $\pm$ 4.9             & 22.9 $\pm$ 3.2   \\
\hline
Total background             & 78.6 $\pm$ 15.2            & 58.4 $\pm$ 11.0            & 82.4 $\pm$ 13.1            & 55.9 $\pm$ 10.2  \\
\hline
Signal + background          & 96.2 $\pm$ 15.3            & 69.6 $\pm$ 11.0            & 100.0 $\pm$ 13.2           & 66.6 $\pm$ 10.2  \\
\hline
Data                         & 112                        & 72                         & 139                        & 82   \\
\hline
\end{tabular}
\end{center}
\label{tab:evtYield_with_datadriven}
\end{table}

\section{The analyses}

  \newcommand{\mt}{\ensuremath{M_{l\nu b}}\xspace}
  \newcommand{\cosThetaPol}{\ensuremath{\cos{\theta^*_{lj}}}\xspace}
  \newcommand{\etalj}{\ensuremath{\eta_{lj}}\xspace}

\subsection{The 2D Analysis}
\label{sec:2d}

\begin{figure}[t]
\centering
\includegraphics[width=0.48\textwidth]{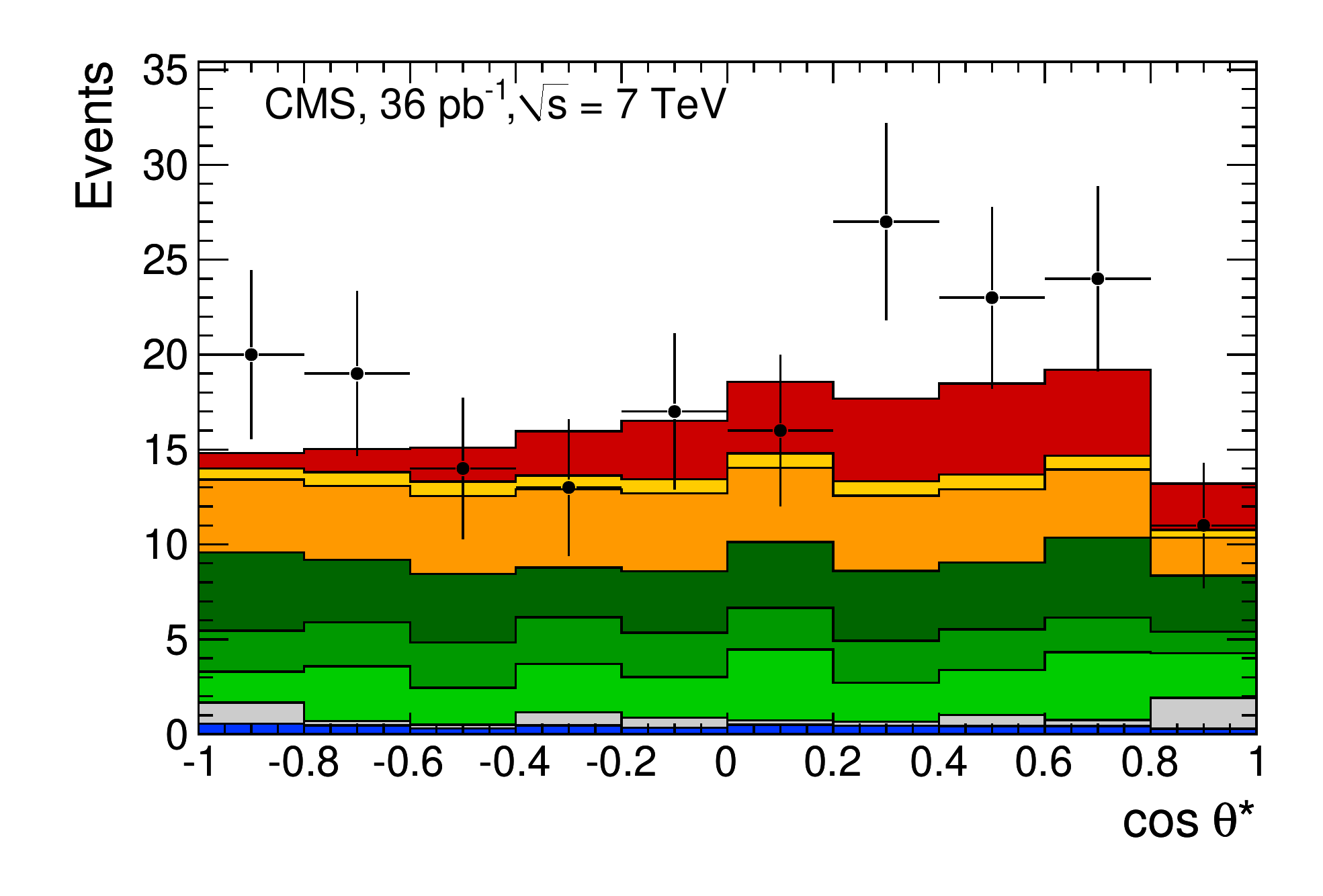}
\includegraphics[width=0.48\textwidth]{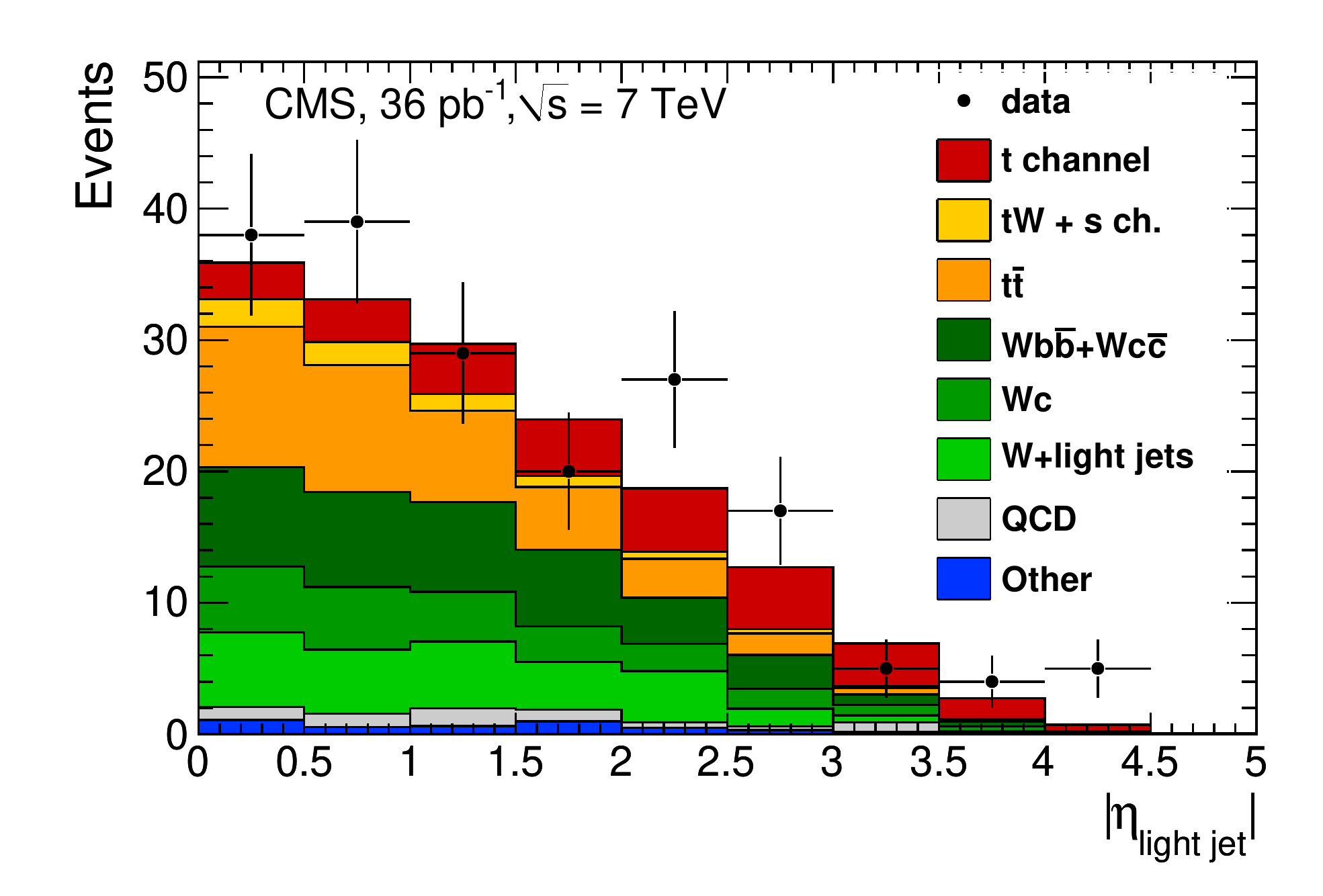}
\caption{Cosine of the angle between charged lepton and untagged jet in the reconstructed top rest frame (\cosThetaPol), left, and pseudorapidity of the untagged jet (\etalj), right, after the full event selection, combining the muon and electron decay channels.
The dip at \cosThetaPol~$\approx 1$ is due to the lepton $p_T$ and $M_T$ selection cuts.
}
\label{fig:2d}
\end{figure}

The cross section is determined by performing an unbinned likelihood fit
 to the 2D distribution of two variables, \cosThetaPol and \etalj.
The distributions of these two variables are shown in Fig.~\ref{fig:2d}.

A property specific to single top production is the almost 100\% left-hand 
 polarization of the top quark due to the $V-A$ structure of the
 electroweak interaction~\cite{Mahlon:1996pn,Motylinski:2009kt}.
Because the lifetime of the top quark is shorter than the hadronization
 scale, the direction of the top-quark spin is visible in angular correlations of its decay products.
 These  are distributed according to
\begin{equation}
\frac{1}{\Gamma}\frac{d\Gamma}{d\cos{\theta^*_{lj}}} = \frac{1}{2}(1+A\cos{\theta^*_{lj}})~,
\label{eq:polarization}
\end{equation}
where $\theta^*_{lj}$ is the angle between the direction of the outgoing lepton and the spin axis, approximated by the direction of the untagged jet, in the top-quark rest frame. $A$ is the coefficient of spin asymmetry, equal to $+1$ for charged leptons.

Another important feature of the signal is the presence of a recoil jet, from the fragmentation of a light (untagged) quark, with a characteristic $\eta$ distribution.

The inputs to the fit are the template distributions for signal and backgrounds, with separate templates for each lepton. 
For the backgrounds, a $\approx 2\%$ correlation is neglected and  the 2D distribution is taken as the product of the 1D templates. The shapes of the discriminating variables for the QCD and $W+$light partons components are taken from the control samples, while all other shapes are taken from simulation. The overall background floats unconstrained in the fit, while its relative components are fixed according to the background estimates. 


\subsection{The BDT Analysis}
\label{sec:bdt}

This analysis assesses the compatibility of the data with the Standard Model predictions of electroweak top quark production using a multivariate analysis method.
Boosted Decision Trees (BDT) are used, with $1000$ decision trees and the
 ADA boosting algorithm as implemented in the TMVA package~\cite{TMVA2007}.

\begin{figure}[t]
\centering
\includegraphics[width=0.48\textwidth]{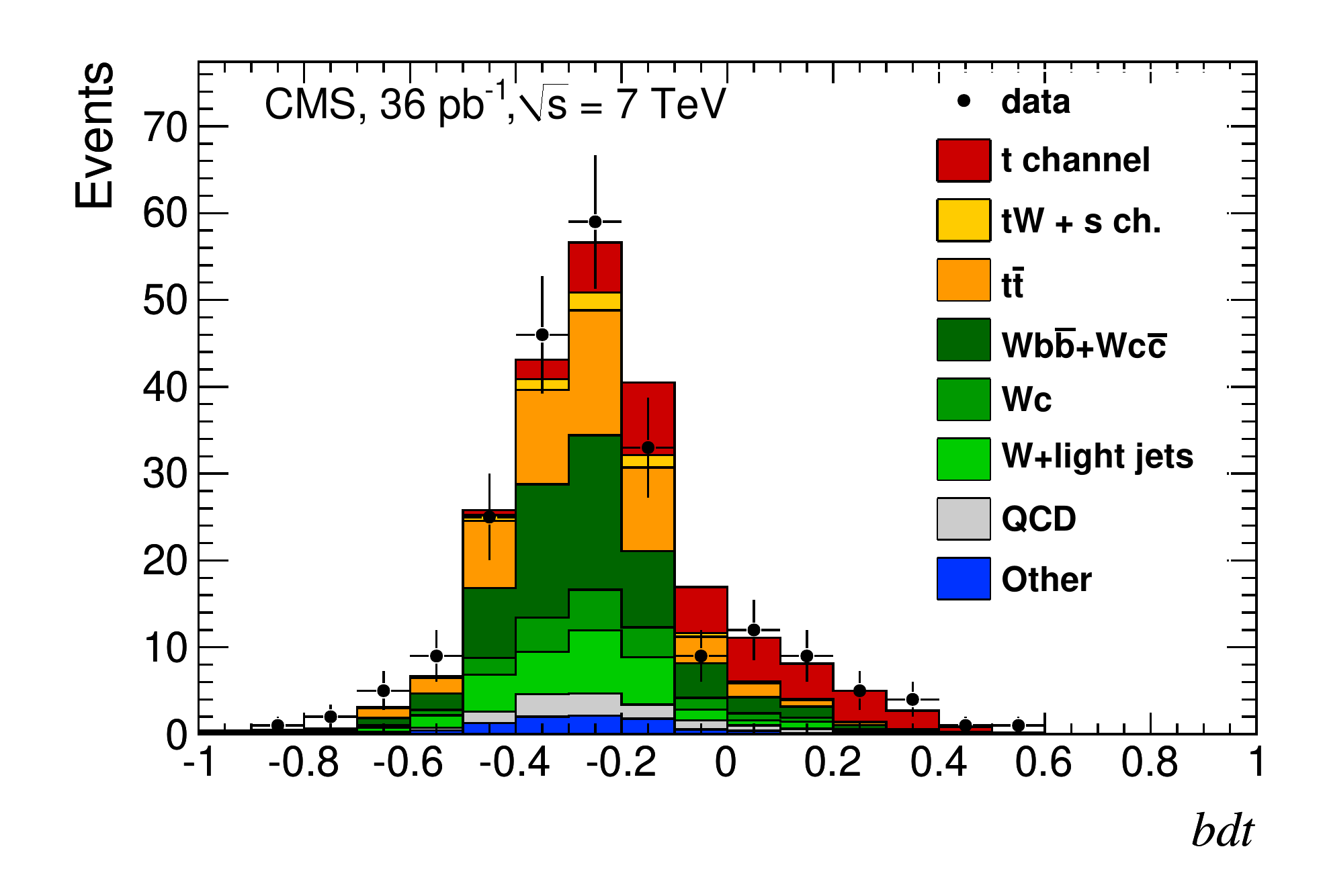}
\caption{Boosted decision tree discriminant ($bdt$) after the dedicated BDT selection, combining the muon and electron decay channels. Predicted backgrounds are scaled to the medians of their posteriors from the fit.}
\label{fig:bdt_e_mu}
\end{figure}

A total of 37 observables reconstructed in the detector have been chosen from five categories. 
The validity of the description of the input variables in the simulation
 has been checked using a Kolmogorov-Smirnov test in the orthogonal
 $W$-enriched control sample. 
The first type of observables covers the kinematics and properties of the leptons and the jets (this includes \etalj), while the second type refers to correlations between these objects.
A third type results from properties of their combinations, the W-boson,
 the top quark, and the sum of the hadronic four-momenta.
A fourth type of observables, which includes \cosThetaPol, exploits the angular distributions between original (lepton, jet) and derived objects ($W$, top quark, etc.).
A fifth type are the event related observables, such as the sphericity and
 the total and transverse energies contained in the parton collision
 process. 
In all these observables, the description of the measured distributions by
 the simulated data is found to be reasonable within the theoretical
 uncertainties.
The most important observables are the lepton momentum, $\hat s$ defined as
 the mass of the system formed by the reconstructed $W$ boson and the two
 jets, the \pt\ of the system formed by the two jets, the \pt\ of the most
 $b$-tagged jet, and the reconstructed top mass.
The $bdt$ classifier has been validated both in simulation and in data. It is shown in Fig.~\ref{fig:bdt_e_mu}.

The cross section is then extracted from a binned likelihood fit to the $bdt$ distribution with a Bayesian approach, where the normalizations of all backgrounds except the multi-jet QCD background and the other systematic uncertainties are treated as nuisance parameters in the fit. For the multi-jet QCD background, the data-driven estimate is used.

\section{Measurement of the production cross section}
\begin{table}
\begin{center}
\begin{tabular}{ |l||c|c|}
\hline
Analysis, channel     & expected           & observed  \\
\hline \hline
2D, $\mu$-channel     & 1.7$^{+1.1}_{-1.0}$ & 2.5 \\     
2D, $e$-channel       & 1.3$^{+1.0}_{-1.1}$ & 3.1 \\
2D, combined          & 2.1$^{+1.0}_{-1.1}$ & 3.7 \\ 
\hline
BDT, $\mu$-channel    & 2.4$^{+0.9}_{-1.0}$ & 3.1 \\     
BDT, $e$-channel      & 2.0$\pm$1.0        & 1.9 \\
BDT, combined         & 2.9$^{+1.0}_{-0.9}$ & 3.5 \\ 
\hline
\end{tabular}
\end{center}
\caption{Expected and observed significances, in number of Gaussian standard deviations, estimated from pseudo-experiments. The uncertainty on the expected significances represents the central 68\% quantile.}
\label{tab:significance}
\end{table}

The 2D analysis yields the following cross section measurements:
\begin{eqnarray}
\sigma^{2D} =  104.1 \pm 42.3 ~\mathrm{(stat.)}  \, _{-28.0}^{+24.8} ~\mathrm{(syst.)}   \pm 4.2 ~\mathrm{(lumi.)}  \,\, \text{pb} & \quad \text{muon channel}  \\
\label{eq:xsec_2d_e}
\sigma^{2D} = 154.2 \pm 56.0 ~\mathrm{(stat.)}  \, _{-46.6}^{+40.6} ~\mathrm{(syst.)}   \pm 6.2 ~\mathrm{(lumi.)}  \,\, \text{pb} & \quad \text{electron channel}  \\
\label{eq:xsec_2d}
\sigma^{2D} = 124.2 \pm 33.8 ~\mathrm{(stat.)}  \, _{-33.9}^{+30.0} ~\mathrm{(syst.)}   \pm 5.0 ~\mathrm{(lumi.)}  \,\, \text{pb} & \quad \text{combined}
\end{eqnarray}
 When combining the electron and the muon decay channels, all systematic uncertainties are considered fully correlated with exception of the data-driven uncertainty on multi-jet QCD.

In the BDT analysis, the following cross sections are measured:
\begin{eqnarray}
\sigma^{BDT} = 90.4 \pm 35.1 ~\mathrm{(stat.)}  \, _{-19.7}^{+16.5} ~\mathrm{(syst.)}   \pm 3.6 ~\mathrm{(lumi.)}  \,\, \text{pb} & \quad \text{muon channel}  \\
\sigma^{BDT} = 59.2 \pm 35.1 ~\mathrm{(stat.)}  \, _{-13.7}^{+13.1} ~\mathrm{(syst.)}   \pm 2.4 ~\mathrm{(lumi.)}  \,\, \text{pb} & \quad \text{electron channel}  \\
\label{eq:xsec_bdt}
\sigma^{BDT} = 78.7 \pm 25.4 ~\mathrm{(stat.)}  \, _{-14.6}^{+13.2} ~\mathrm{(syst.)}   \pm 3.1 ~\mathrm{(lumi.)}  \,\, \text{pb} & \quad \text{combined}
\end{eqnarray}

The main systematic uncertainties are the uncertainty on b-tagging, the jet energy scale and the modeling of the signal and backgrounds.
The expected and observed significance when including all systematic uncertainties are given in Table~\ref{tab:significance}.
The measurements are consistent among them and with the standard model expectation in the 4- and 5-flavour schemes.
 Both confirm the Tevatron observation of the electroweak mode of top quark production.

The measurements from the 2D and BDT analyses are then combined with the
 Best Linear Unbiased Estimation (BLUE) method~\cite{Lyons1988110}. 
The statistical correlations estimated from simulated samples is 0.51.
Systematic uncertainties common to both methods are assumed to be 100\% correlated.
The combined result is:
$$
\sigma = 83.6 \pm 29.8~\mathrm{(stat.+syst.)} \pm 3.3~\mathrm{(lumi.)\,\,pb}
$$

This result can be used to derive an estimate of CKM matrix element $|V_{tb}|$.
With the assumption that $|V_{td}|$ and $|V_{ts}|$ are much smaller than $|V_{tb}|$ and using the NLO prediction in the 5-flavors scheme $\sigma^{th}=62.3^{+2.3}_{-2.4}$~pb~\cite{Campbell:2009gj}, $|V_{tb}|$ is found to be
$$
|V_{tb}| = \sqrt{\frac{\sigma^{exp}}{\sigma^{th}}} = 1.16 \pm 0.22~\mathrm{(exp.)} \pm 0.02 ~\mathrm{(th.)}
$$

\section{Conclusion}

A first measurement of the production cross section of single top quark  pp collisions at $\sqrt{s} = 7$~TeV  was performed on an integrated luminosity of 36~pb$^{-1}$ recorded at CMS.
Two separate analyses were made, and the  combination of the two measurements yields  $\sigma = 83.6 \pm 29.8~\mathrm{(stat.+syst.)} \pm 3.3~\mathrm{(lumi.)\,\,pb}$. This measurement is consistent with the SM prediction.

\bigskip 

\end{document}